\documentclass{elsarticle}
\usepackage{amsmath,amssymb}
\usepackage{epsfig,graphicx}
\usepackage{color}

\begin{document}
\title{\bf  Tunneling cosmological state revisited: Origin of inflation
with a non-minimally coupled Standard Model Higgs inflaton}
\author[LPI]{Andrei O. Barvinsky}
\ead{barvin@td.lpi.ru}

\author[BU,LI]{Alexander Yu. Kamenshchik}
\ead{kamenshchik@bo.infn.it}

\author[KU]{Claus Kiefer}
\ead{kiefer@thp.uni-koeln.de}

\author[KU]{Christian F. Steinwachs}
\ead{cst@thp.uni-koeln.de}

\address[LPI]{Theory Department, Lebedev
Physics Institute,
Leninsky Prospect 53, Moscow 119991, Russia}
\address[BU]{Dipartimento di Fisica and INFN,
via Irnerio 46, 40126 Bologna, Italy}
\address[LI]{L.D. Landau Institute for
Theoretical Physics, Moscow 119334, Russia}
\address[KU]{Institut f\"ur Theoretische Physik,
Universit\"at zu K\"oln, Z\"ulpicher Strasse 77,
50937 K\"oln, Germany}

\begin{abstract}
We suggest a path integral formulation for the tunneling
cosmological state, which admits a consistent renormalization and
renormalization group (RG) improvement in particle physics
applications of quantum cosmology. We apply this formulation to the
inflationary cosmology driven by the Standard Model (SM) Higgs boson
playing the role of an inflaton with a strong non-minimal coupling
to gravity. In this way a complete cosmological scenario is
obtained, which embraces the formation of initial conditions for the
inflationary background in the form of a sharp probability peak in
the distribution of the inflaton field and the ongoing generation of
the Cosmic Microwave Background (CMB)  spectrum on this
background. Formation of this probability
peak is based on the same RG mechanism which underlies the
generation of the CMB spectrum which was recently shown to be
compatible with the WMAP data in the Higgs mass range $135.6\; {\rm
GeV} \lesssim M_H\lesssim 184.5\;{\rm GeV}$. This brings to life a
convincing unification of quantum cosmology with the particle
phenomenology of the SM, inflation theory, and CMB observations.
\end{abstract}


\maketitle
\section{Introduction}

At the dawn of inflation theory two prescriptions for the quantum
state of the Universe were seriously considered as a source of
initial conditions for inflation. These are the so-called
no-boundary \cite{noboundary} and tunneling \cite{tunnel,Vilenkin84}
cosmological wavefunctions (see also \cite{OUP} for a general review),
whose semiclassical amplitudes are
roughly inversely proportional to one another. In the model of chaotic
inflation driven in the slow-roll approximation by the inflaton
field $\varphi$ with the potential $V(\varphi)$ these amplitudes
read as $|\varPsi_\pm(\varphi)|\simeq\exp(\mp S_E(\varphi)/2)$,
where $+/-$ label, respectively, the no-boundary/tunneling
wavefunctions. Here, $S_E(\varphi)$ is the Einstein action of the
Euclidean de Sitter instanton $S^4$ with the effective cosmological
constant given by the value of the inflaton field $\Lambda_{\rm
eff}=V(\varphi)/M_{\rm P}^2$,
    \begin{eqnarray}
    S_E(\varphi)\simeq
    -\frac{24\pi^2 M_{\rm P}^4}{V(\varphi)},          \label{S_E}
    \end{eqnarray}
in units of the reduced Planck mass $M_{\rm P}^2=1/8\pi G$
($\hbar=1=c$). The
no-boundary state was originally formulated as a path integral over
Euclidean four-geometries; the tunneling state in the form of a path
integral over Lorentzian metrics was presented in
\cite{Vilenkin84,Vilenkin94}, and both wavefunctions were also
obtained as solutions of the minisuperspace Wheeler--DeWitt equation.

The no-boundary and tunneling states lead to opposite physical
conclusions. In particular, in view of the negative value of the
Euclidean de Sitter action the no-boundary state strongly enhances
the contribution of empty universes with $V(\varphi)=0$ in the full
quantum state and, thus, leads to the very counterintuitive
conclusion that infinitely large universes are infinitely more
probable than those of a finite size -- a property which underlies
the once very popular but now nearly forgotten big-fix mechanism of S.
Coleman \cite{bigfix}. On the other hand, the tunneling state favors
big values of $V(\varphi)$ capable of generating inflationary
scenarios. Thus, it would seem that the tunneling prescription is
physically more preferable than the no-boundary one. However, the
status of the tunneling prescription turns out to be not so simple
and even controversial.

Naive attempts to go beyond the minisuperspace approximation
lead to unnormalizable states in the sector of spatially inhomogeneous
degrees of freedom for matter and metric and invalidate, in particular,
the usual Wick rotation from the Lorentzian to the
Euclidean spacetime. This problem was partly overcome by imposing
the normalizability condition on the matter part of the solution of
the Wheeler--DeWitt equation \cite{Vilenkin88}, but the situation
remained controversial for the following reason.

Modulo the issue of quantum interference between the ``contracting''
and ``expanding'' branches of the cosmological wavefunction
discussed, for example, in \cite{Vilenkin88,OUP,CK92,debate}, the
amplitudes of the no-boundary and tunneling branches of such a
semiclassical solution take the form
    \begin{eqnarray}
    \left|\,\varPsi_\pm
    \big(\varphi,\varPhi({\bf x})\big)\right|
    =\exp\left({\mp\frac12S_E(\varphi)}\right)
    \left|\,\varPsi_{\rm matter}
    \big(\varphi,\varPhi({\bf x})\big)\right|,
    \end{eqnarray}
where $\varPhi({\bf x})$ is a set of matter fields separate from
the spatially homogeneous inflaton, and $\varPsi_{\rm matter}
\big(\varphi,\varPhi({\bf x})\big)$ is their normalizable
(quasi-Gaussian) part in the full wavefunction -- in essence
representing the Euclidean de Sitter invariant vacuum of linearized
fields $\varPhi({\bf x})$ on the quasi-de Sitter background with
$\Lambda_{\rm eff}=V(\varphi)/M_{\rm P}^2$. Quantum averaging over
$\varPhi({\bf x})$ then leads to the following quantum distribution of
the inflaton field
    \begin{eqnarray}
    \rho_\pm^{\rm 1-loop}(\varphi)=
    \int d\big[\,\varPhi({\bf x})\,\big]\left|\,\varPsi_\pm
    \big(\varphi,\varPhi({\bf x})\big)\right|^2
    =\exp\left(\mp S_E(\varphi)
    -S_E^{\rm 1-loop}(\varphi)\right)\ ,    \label{1000}
    \end{eqnarray}
where $S_E^{\rm 1-loop}(\varphi)=(1/2){\rm Tr}
\ln(\delta^2S_E[\,\varphi,\varPhi\,]/\delta\varPhi(x)\,\delta\varPhi(y))$
is the contribution of the UV divergent one-loop effective action
\cite{norm,PhysRep,reduc}. With the aid of this algorithm a sharp
probability peak was obtained in the {\em tunneling} distribution
$\rho_-^{\rm 1-loop}(\varphi)$ for the model with a strong
non-minimal coupling of the inflaton to gravity
\cite{norm,we-scale,BK}. This peak was interpreted as generating the
quantum scale of inflation -- the initial condition for its
inflationary scenario. Quite remarkably, for accidental reasons this
result was free from the usual UV renormalization ambiguity. It did
not require application of the renormalization scheme of absorbing
the UV divergences into the redefinition of the coupling constants
in the tree-level action $S_E(\varphi)$.

However, beyond the one-loop approximation and for other physical
correlators the situation changes, and one has to implement a UV
renormalization in full. But with the $\mp S_E(\varphi)$ ambiguity
in (\ref{1000}) this renormalization would be different for the
tunneling and no-boundary states. For instance, an asymptotically free theory
in the no-boundary case (associated with the usual Wick rotation to
the Euclidean spacetime) will not be asymptotically free in the
tunneling case. The tunneling versus no-boundary gravitational
modification of the theory will contradict basic field-theoretical
results in flat spacetime. This strongly invalidates a naive
construction of the tunneling state of the above type. In
particular, it does not allow one to go beyond the one-loop
approximation in the model of non-minimally coupled inflaton and
perform its renormalization group (RG) improvement.

Here we suggest a solution of this problem by formulating a new path
integral prescription for the tunneling state of the Universe. This
formulation is based on a recently suggested construction of the
cosmological density matrix \cite{slih} which describes a
microcanonical ensemble of cosmological models \cite{why}. The
statistical sum of this ensemble was calculated in a spatially
closed model with a generic set of scalar, spinor, and vector fields
conformally coupled to gravity. It was obtained in the saddle-point
approximation dominated by the contribution of the thermal
cosmological instantons of topology $S^3\times S^1$. These
instantons also include the vacuum $S^4$ topology treated as a
limiting case of the compactified time dimension $S^1$ in $S^3\times
S^1$ being ripped in the transition from $S^3\times S^1$ to $S^4$.
This limiting case exactly recovers the Hartle--Hawking state of
\cite{noboundary}, so that the whole construction of \cite{slih,why}
can be considered as a generalization of the vacuum no-boundary
state to the quasi-thermal no-boundary ensemble. The basic physical
conclusion for this ensemble was that it exists in a bounded range
of values of the effective cosmological constant, that it is capable
of generating a big-boost scenario of the cosmological acceleration
\cite{bigboost} and that its vacuum Hartle--Hawking member does not
really contribute because it is suppressed by the infinite {\em
positive} value of its action. This is a genuine effect of the
conformal anomaly of quantum fields \cite{FHH,Starobinsky}, which
qualitatively changes the tree-level action (\ref{S_E}).

Below we shall show that the above path integral actually has another
saddle point corresponding to the negative value of the lapse
function $N<0$, which is gauge-inequivalent to $N>0$. In the main,
this leads to the inversion of the sign of the action in the
exponential of the statistical sum and, therefore, deserves the label
``tunneling''. In this tunneling state the thermal part vanishes
and its instanton turns out to be a purely vacuum one. Finally, this
construction no longer suffers from the above mentioned controversy
with the renormalization. A full quantum effective action is
supposed to be calculated and renormalized by the usual set of
counterterms on the background of a generic metric and then the
result should be analytically continued to $N<0$ and taken at the
{\em tunneling} saddle point of the path integral over the lapse
function $N$.

Below we shall apply this construction to a cosmological model for which the
Lagrangian of the graviton-inflaton sector reads
    \begin{eqnarray}
    &&{\mbox{\boldmath $L$}}(g_{\mu\nu},\varPhi)=
    \frac12\left(M_{\rm P}^2+\xi|\varPhi|^2\right)R
    -\frac{1}{2}|\nabla\varPhi|^{2}
    -V(|\Phi|),                         \label{inf-grav}\\
    &&V(|\Phi|)=
    \frac{\lambda}{4}(|\varPhi|^2-v^2)^2,\,\,\,\,
    |\varPhi|^2=\varPhi^\dag\varPhi,
    \end{eqnarray}
where $\varPhi$ is the Standard Model (SM) Higgs boson, whose
expectation value plays the role of an inflaton and which is assumed
here to possess a
strong non-minimal curvature coupling with $\xi\gg 1$. Here, as
above, $M_{\rm P}$ is a reduced Planck mass, $\lambda$ is a quartic
self-coupling of $\varPhi$, and $v$ is an electroweak (EW) symmetry
breaking scale.

The early motivation for this model with a GUT type boson $\varPhi$
\cite{non-min,KomatsuFutamase} was to avoid an exceedingly small
quartic coupling $\lambda$ by invoking a non-minimal coupling with a
large $\xi$. This was later substantiated by the hope to generate
the no-boundary/tunneling initial conditions for inflation
\cite{we-scale,BK}. This theory but with the SM Higgs
boson $\varPhi$ instead of the abstract GUT setup of
\cite{we-scale,BK} was suggested in \cite{BezShap}, extended in
\cite{we} to the one-loop level and considered regarding its
reheating mechanism in \cite{GB08}. The RG
improvement in this model has predicted CMB parameters -- the
amplitude of the power spectrum and its spectral index -- compatible
with WMAP observations in a finite range of values of the Higgs
mass, which is close to the widely accepted range dictated by the EW
vacuum stability and perturbation theory bounds
\cite{Wil,BezShap1,BezShap3,RGH,PLBRGH,Clarcketal}.

The purpose of our paper is to extend
the results of \cite{RGH,PLBRGH} by suggesting that this model
does not only have WMAP-compatible CMB perturbations, but can also generate
the initial conditions for the inflationary background upon which
these perturbations propagate. These initial conditions are realized
in the form of a {\em sharp probability peak} in the tunneling
distribution function of the inflaton.

Our paper is organized as follows. In Sect.~2 we present
the path-integral formulation for the tunneling state and derive the
relevant distribution in the space of values of the cosmological
constant. In Sect.~3 we apply this distribution to the gravitating SM
model with the graviton-inflaton sector (\ref{inf-grav}) and obtain
the probability peak in the distribution of the initial value of the
Higgs-inflaton. Sect.~4 contains a short discussion.

%%%%%%%%%%%%%%%%%%%%%%%%%%%%%%%%%%%%%%%%%%%%%%%%%%%%%%%%%%%%%%%

\section{Tunneling cosmological wavefunction within the path
integral formulation}

The path integral for the microcanonical statistical sum in
cosmology \cite{why} can be cast into the form of an integral over
a minisuperspace lapse function $N(\tau)$ and scale factor
$a(\tau)$ of a spatially closed Euclidean FRW metric $ds^2 =
N^2(\tau)\,d\tau^2 +a^2(\tau)\,d^2\Omega^{(3)}$,
    \begin{eqnarray}
    &&e^{-\varGamma}=\int
     D[\,a,N\,]\;
    e^{-S_{\rm eff}[\,a,\,N\,]},              \label{1}\\
    &&e^{-S_{\rm eff}[\,a,\,N]}
    =\int D\varPhi(x)\,
    e^{-S[\,a,\,N;\,\varPhi(x)\,]}\ .             \label{2}
    \end{eqnarray}
Here, $S_{\rm eff}[\,a,\,N\,]$ is the Euclidean effective action of
all inhomogeneous ``matter'' fields
$\varPhi(x)=(\phi(x),\psi(x),A_\mu(x), h_{\mu\nu}(x),...)$ (which
include also metric perturbations) on the minisuperspace background
of the FRW metric, $S[a,N;\varPhi(x)]$ is the classical Euclidean
action, and the integration runs over periodic fields on the
Euclidean spacetime with a compactified time $\tau$ (of $S^1\times
S^3$ topology).

It is important that the integration over the lapse function $N$
runs along the imaginary axis from $-{\rm i}\infty$ to $+{\rm i}\infty$ because
this Euclidean path integral represents, in fact, the transformed
version of the integral over metrics with Lorentzian signature.
This transformation is the usual Wick rotation which can be
incorporated by the transition from the Lorentzian lapse function
$N_{\rm L}$ to the Euclidean one $N$ by the relation $N_{\rm L}={\rm
  i}N$ \cite{why}.
The Lorentzian path integral, in turn, fundamentally follows
from the definition of the microcanonical ensemble in quantum
cosmology which includes all {\em true} physical configurations
satisfying the quantum first-class constraints -- the Wheeler--DeWitt
equations. The projector onto these configurations is realized in
the integrand of the path integral by the delta functions of the
Hamiltonian (and momentum) constraints. The Fourier representation
of these delta functions in terms of the integral over the
conjugated Lagrange multipliers -- the lapse $N_{\rm L}$ (and shift)
functions -- implies an integration with limits at infinity,
$-\infty<N_{\rm L}<\infty$, which explains the range of integration over
the Euclidean $N$.

It should be mentioned that a full non-perturbative evaluation of the
path integral would require a careful inspection of the infinite
contours in the complex $N$-plane that render the integral convergent,
see, for example, \cite{pathintegrals}. However, such an inspection is
not needed here because we are dealing here with a semiclassical
approximation in which only the vicinity of the saddle point enters.

The convenience of writing the path integral (\ref{1}) in the
Euclidean form follows from the needs of the semiclassical
approximation. In this approximation, it is dominated by the
contribution of a saddle point, $\varGamma_0=S_{\rm
eff}[\,a_0,N_0\,]$, where $a_0=a_0(\tau)$ and $N_0=N_0(\tau)$
solve the equation of motion for $S_{\rm eff}[\,a,N\,]$ and satisfy
periodicity conditions dictated by the definition of the statistical
sum. Such periodic solutions exist in the Euclidean domain
with real $N$ rather than in the Lorentzian one with the imaginary
lapse. This means that the contour of integration over $N$ along the
imaginary axis should be deformed into the complex plane to traverse
the real axis at some $N_0\neq 0$ corresponding to the Euclidean
solution of the equations of motion for the minisuperspace action.

The residual one-dimensional diffeomorphism invariance of this
action (which is gauged out by the gauge-fixing procedure implicit
in the integration measure $D[\,a,N\,]$) allows one to fix the
ambiguity in the choice of $N_0$. There remains only a double-fold
freedom in this choice actually inherited from the sign
indefiniteness of the integration range for $N_L$. This freedom is
exhausted by either positive, $N_0>0$, or negative, $N_0<0$, values
of the lapse, because, on the one hand, all values in each of these
equivalence classes are gauge equivalent and, on the other hand, no
continuous family of nondegenerate diffeomorphisms exists relating
these classes to one another. Without loss of generality one can
choose as representatives of these classes $N_\pm=\pm 1$ and label
the relevant solutions and on-shell actions, respectively, as
$a_\pm(\tau)$ and
    \begin{eqnarray}
    \varGamma_\pm=S_{\rm eff}[\,a_\pm(\tau),\pm 1\,]\ . \label{pm}
    \end{eqnarray}
Gauge inequivalence of these two cases, $\varGamma_-\neq
\varGamma_+$, is obvious because, for example, all local
contributions to the effective action are odd functionals of $N$,
$S_{\rm local}[\,a,N\,]=-S_{\rm local}[\,a,-N\,]$. Thus we
can heuristically identify
the statistical sums $\varGamma_\pm$ correspondingly with the
``no-boundary'' and ``tunneling'' prescriptions for the quantum state
of the Universe,
    \begin{eqnarray}
    \exp(-\varGamma_{\rm no-boundary/tunnel})=e^{-\varGamma_\pm}.
    \end{eqnarray}
In other words, we use this equation to define ``no-boundary'' and
``tunneling'' in the first place. This result shows that for both
prescriptions a full quantum effective action as a whole sits in the
exponential of the partition function without any splitting into the
minisuperspace and matter contributions weighted by different sign
factors like in (\ref{1000}). This means that the usual
renormalization scheme is applicable to the calculation of
(\ref{pm}) -- generally covariant UV counterterms should be
calculated on the background of a generic metric and afterwards
evaluated at the FRW metric with $N=\pm 1$, depending on the choice
of either the no-boundary or tunneling prescription. Below we
demonstrate how this procedure works for the system dominated by
quantum fields with heavy masses, whose effective action admits a
local expansion in powers of the spacetime curvature and matter
fields gradients.

For such a system the Euclidean effective action takes the form
    \begin{equation}
    S_{\rm eff}[g_{\mu\nu}]=\int d^{4}x\,g^{1/2}
    \left(M_{\rm P}^2\Lambda-\frac{M_{\rm P}^2}2\,R(g_{\mu\nu})
    +...\right),                               \label{effaction0}
    \end{equation}
where we disregard the terms of higher orders in the curvature and
derivatives of the mean values of matter fields. Here the
cosmological term and the (reduced) Planck mass squared
$M_{\rm P}^2=1/8\pi G$ can be considered as functions of these mean values
and treated as constants in the approximation of slowly varying
fields. This effective action does not contain the thermal part
characteristic of the statistical ensemble \cite{slih} because for
heavy quanta the radiation bath is not excited. This is justified by
the fact that the effective temperature of this bath turns out to be
vanishing.

In fact, the minisuperspace action functional for (\ref{effaction0})
reads in units of $m_{\rm P}^2=3\pi/4G=6\pi^2M_{\rm P}^2$ as
    \begin{eqnarray}
    S_{\rm eff}[\,a,N\,]
    =m_{\rm P}^2\int d\tau\,N (-aa'^2
    -a+ H^2 a^3),              \label{Seff}
    \end{eqnarray}
where $a'\equiv da/Nd\tau$, and we use the notation for the
cosmological constant $\Lambda=3H^2$ in terms of the effective
Hubble factor $H$. Then the saddle point for the path integral
(\ref{1}) -- the stationary configuration with respect to variations
of the lapse function, $\delta S_{\rm eff}[\,a,N\,]/\delta N=0$, --
satisfies the Euclidean Friedmann equation
    \begin{eqnarray}
    a'^2=1-H^2a^2.              \label{eq}
    \end{eqnarray}
It has one turning point at $a_+=1/H$ below which the real solution
interpolates between $a_-\equiv a(0)=0$ and $a_+$. In the gauge
$N=\pm 1$ for both no-boundary/tunneling cases this solution
describes the Euclidean de Sitter metric, that is, one hemisphere of $S^4$,
    \begin{eqnarray}
    a_\pm(\tau)=\frac1H\sin(H\tau).    \label{dS}
    \end{eqnarray}
After the bounce from the equatorial section of the maximal scale
factor $a_+$, this solution spans at the contraction phase the rest
of the full four-sphere\footnote{The formal analytic extension from
$N_0=1$ to $N_0=-1$ should not, of course, be applied to
$a(\tau)=\sin(N_0H\tau)/H$ to give a negative $a(\tau)$ instead of
(\ref{dS}), because in contrast to the sign-indefinite Lagrange
multiplier $N$ the path integration over $a(\tau)$ in (\ref{1})
semiclassically always runs in the vicinity of its positive
geometrically meaningful value. For this reason, $a(\tau)$ never
brings sign factors into the on-shell action even though it enters the
action with odd powers.}. Thus, this solution is not periodic and in
the terminology of \cite{slih} describes a purely vacuum
contribution to the statistical sum (\ref{1}). As shown in
\cite{slih}, the effective temperature of this state is determined
by the inverse of the full period of the instanton solution measured
in units of the conformal time $\eta$. Therefore, for (\ref{dS}) it
vanishes because this period between the poles of this spherical
instanton is divergent,
    \begin{eqnarray}
    \eta=2\int_0^{\pi/2H} \frac{d\tau}{a(\tau)}\to\infty\ .
    \end{eqnarray}
This justifies the absence of the thermal part in
(\ref{effaction0}).

Thus, with $N=\pm 1$ the no-boundary and tunneling on-shell actions
(\ref{pm}) read
    \begin{eqnarray}
    \varGamma_\pm=\mp\frac{8\pi^2M_{\rm P}^2}{H^2}
    \end{eqnarray}
and the object of major interest here -- the tunneling partition
function in the space of positive values of $H^2=\Lambda/3$ -- is
given by
    \begin{eqnarray}
    \rho_{\rm tunnel}(\Lambda)=
    \exp\left(-\frac{24\pi^2M_{\rm P}^2}{\Lambda}\right),
    \,\,\,\,\,\Lambda>0.    \label{partitionH}
    \end{eqnarray}
It coincides with the semiclassical tunneling wavefunction of the
Universe \cite{tunnel}, $|\varPsi_{\rm
tunnel}|^2\simeq\exp(-8\pi^2M_{\rm P}^2/H^2)$, derived from the
Wheeler--DeWitt equation in the tree-level approximation.

At the turning point $a_+$, the solution (\ref{dS}) can be
analytically continued to the Lorentzian regime,
$a_{\rm L}(t)=a(\pi/2H+{\rm i}t)$. The scale factor then expands
eternally as
    \begin{eqnarray}
    a_{\rm L}(t)=\frac1H\cosh(Ht)\ ,
    \end{eqnarray}
which can be interpreted as representing the distributions of scale factors
in the quantum ensemble (after decoherence) of de Sitter models
distributed according to (\ref{partitionH}). Note that the attempt
to extend this ensemble to negative $\Lambda$ fails, because the
equation (\ref{eq}) with $H^2<0$ does not have turning points with
nucleating real Lorentzian geometries. Moreover, virtual
cosmological models with Euclidean signature are also forbidden in
the tunneling state because their positive Euclidean action diverges
to infinity, so that $\rho_{\rm tunnel}(\Lambda)=0$ for $\Lambda<0$.

%%%%%%%%%%%%%%%%%%%%%%%%%%%%%%%%%%%%%%%%%%%%%%%%%%%%%%%%%%%%%%%%%%%%

\section{Quantum origin of the Universe with the SM Higgs-inflaton
non-minimally coupled to curvature}

The partition function of the above type can serve as a source of
initial conditions for inflation only when the cosmological constant
$\Lambda=3H^2$ becomes a composite field capable of a decay at the
exit from inflation. Usually this is a scalar inflaton field whose
quantum mean value $\varphi$ is nearly constant in the slow roll
regime, and its effective potential $V(\varphi)$ plays the role of
the cosmological constant driving the inflation. When the
contribution of the inflaton gradients is small, the above formalism
remains applicable also with the inclusion of this field whose
ultimate effect reduces to the generation of the effective
cosmological constant $\Lambda=V(\varphi)/M_{\rm P}^2$ and the effective Planck
mass.

These constants are the coefficients of the zeroth and first order
terms of the effective action expanded in powers of the curvature,
and they incorporate radiative corrections due to all quantum fields
in the path integral (\ref{2}). Now there is no mismatch between the
signs of the tree-level and loop parts of the partition function.
Therefore, one can apply the usual renormalization and, if necessary,
the renormalization group (RG) improvement to obtain the full
effective action $S_{\rm eff}[g_{\mu\nu},\varphi]$ and then repeat
the procedure of the previous section. In the slow roll
approximation the effective action has the general form
    \begin{equation}
    S_{\rm eff}[g_{\mu\nu},\varphi]=\int d^{4}x\,g^{1/2}
    \left(V(\varphi)-U(\varphi)\,R(g_{\mu\nu})+
    \frac12\,G(\varphi)\,
    (\nabla\varphi)^2+...\right),   \label{effaction}
    \end{equation}
where $V(\varphi)$, $U(\varphi)$ and $G(\varphi)$ are the
coefficients of the derivative expansion, and we disregard the
contribution of higher-derivative operators. With the slowly varying
inflaton the coefficients $V(\varphi)$ and $U(\varphi)$ play the
role of the effective cosmological and Planck mass constants, so
that one can identify in (\ref{effaction0}) and (\ref{Seff}) the
effective $M_{\rm P}^2=m_{\rm P}^2/6\pi^2$ and $H^2$, respectively, with
$2U(\varphi)$ and $V(\varphi)/6U(\varphi)$. Therefore, the tunneling
partition function (\ref{partitionH}) becomes the following
distribution of the field $\varphi$
    \begin{eqnarray}
    &&\rho_{\rm tunnel}(\varphi)=
    \exp\left(-\frac{24\pi^2M_{\rm P}^4}
    {\hat V(\varphi)}\right),                \label{partitionphi}\\
    &&\hat V(\varphi)=
    \left(\frac{M_{\rm P}^2}{2}\right)^2
    \frac{V(\varphi)}{U^2(\varphi)},        \label{hatV}
    \end{eqnarray}
where $\hat V(\varphi)$ in fact coincides with the potential in the
Einstein frame of the action (\ref{effaction}) \cite{RGH,PLBRGH}.

Now we apply this formalism to the model (\ref{inf-grav}) of
inflation driven by the SM Higgs inflaton
$\varphi=(\varPhi^\dagger\varPhi)^{1/2}$. As shown in
\cite{RGH,PLBRGH}, the one-loop RG improved action in this model has
for large $\varphi$ the form (\ref{effaction}) with the coefficient
functions
    \begin{eqnarray}
    &&V(\varphi)=
    \frac{\lambda(t)}{4}\,Z^4(t)\,\varphi^4,  \label{RGeffpot}\\
    &&U(\varphi)=
    \frac12\Big(M_{\rm P}^2
    +\xi(t)\,Z^2(t)\,\varphi^{2}\Big),      \label{RGeffPlanck}\\
    &&G(\varphi)=Z^2(t),            \label{phirenorm1}
    \end{eqnarray}
determined in terms of the running couplings $\lambda(t)$ and $\xi(t)$,
and the field renormalization $Z(t)$. They incorporate a summation of
powers of logarithms and belong to the solution of the RG equations
which at the inflationary stage with a large $\varphi\sim
M_{\rm P}/\sqrt\xi$ and large $\xi\gg1$
read as (see \cite{RGH,PLBRGH} for details)
    \begin{eqnarray}
    &&\frac{d\lambda}{dt} =
    \frac{{\mbox{\boldmath $A$}}}{16\pi^2}\lambda
    -4\gamma\lambda,                           \label{beta-lambda}\\
    &&\frac{d\xi}{dt} =
    \frac{6\lambda}{16\pi^2}\xi
    -2\gamma\xi                 \label{beta-xi}
    \end{eqnarray}
and $dZ/dt=\gamma Z$. Here, $\gamma$ is the anomalous dimension of
the Higgs field, the running scale $t=\ln(\varphi/M_t)$ is
normalized at the top quark mass $\mu=M_t$, and
$\mbox{\boldmath$A$}=\mbox{\boldmath$A$}(t)$ is the running
parameter of the {\em anomalous scaling}. This quantity was
introduced in \cite{norm} as the pre-logarithm coefficient of the
overall effective potential of all SM physical particles and
Goldstone modes. Due to their quartic, gauge and Yukawa couplings with
$\varphi$, they acquire masses $m(\varphi)\sim\varphi$ and for large
$\varphi$ give rise to the asymptotic behavior of the
Coleman-Weinberg potential,
   \begin{eqnarray}
    &&V^{\rm 1-loop}(\varphi)=\!\!\sum_{
    \rm particles}
    (\pm 1)\,\frac{m^4(\varphi)}{64\pi^2}
    \,\ln\frac{m^2(\varphi)}{\mu^2}
    \simeq\frac{\lambda\mbox{\boldmath$A$}}{128\pi^2}
    \,\varphi^4
    \ln\frac{\varphi^2}{\mu^2},  \label{Aviamasses}
    \end{eqnarray}
which can serve as a definition of $\mbox{\boldmath$A$}$.

The importance of this quantity and its modification due to the RG
running of the non-minimal coupling $\xi(t)$,
        \begin{eqnarray}
        \mbox{\boldmath$A_I$}=
        \mbox{\boldmath$A$}-12\lambda   \label{AI}
        \end{eqnarray}
($\mbox{\boldmath$A_I$}$ gives the running of the ratio
$\lambda/\xi^2$,
$16\pi^2(d/dt)(\lambda/\xi^2)=\mbox{\boldmath$A_I$}(\lambda/\xi^2))$,
is that for $\xi\gg 1$ mainly these parameters determine the quantum
inflationary dynamics \cite{BK,efeqmy} and yield the parameters of
the CMB generated during inflation \cite{we}. In particular, the
value of $\varphi$ at the beginning of the inflationary stage of
duration $N$ in units of the e-folding number turns out to be
\cite{we}
    \begin{eqnarray}
    &&\varphi^2=
    -\frac{64\pi^2M_{\rm P}^2}
    {\xi \mbox{\boldmath$A_I$}(t_{\rm end})}
    (1-e^x),                                  \label{xversusvarphi}\\
    &&x\equiv\frac{N
    \mbox{\boldmath$A_I$}(t_{\rm end})}{48\pi^2},           \label{x}
    \end{eqnarray}
where a parameter $x$ has been introduced which directly involves
$\mbox{\boldmath$A_I$}(t_{\rm end})$ taken at the end of inflation,
$t_{\rm end}=\ln(\varphi_{\rm end}/M_t)$, $\varphi_{\rm end}\simeq
2M_{\rm P}/\sqrt{3\xi}$. This parameter also enters simple algorithms for
the CMB power spectrum $\Delta_\zeta^2(k)$ and its spectral index
$n_s(k)$. As shown in \cite{RGH,PLBRGH}, the application of these
algorithms under the observational constraints
$\Delta_{\zeta}^2(k_0)\simeq 2.5\times 10^{-9}$ and $0.94
<n_s(k_0)<0.99$ (the combined WMAP+BAO+SN data at the pivot point
$k_0=0.002$ Mpc$^{-1}$ corresponding to $N\simeq 60$ \cite{WMAP})
gives the CMB-compatible range of the Higgs mass $135.6\; {\rm GeV}
\lesssim M_{\rm H}\lesssim 184.5\;{\rm GeV}$, both bounds being determined
by the lower bound on the CMB spectral index.

\begin{figure}[h]
\centerline{\epsfxsize 12cm \epsfbox{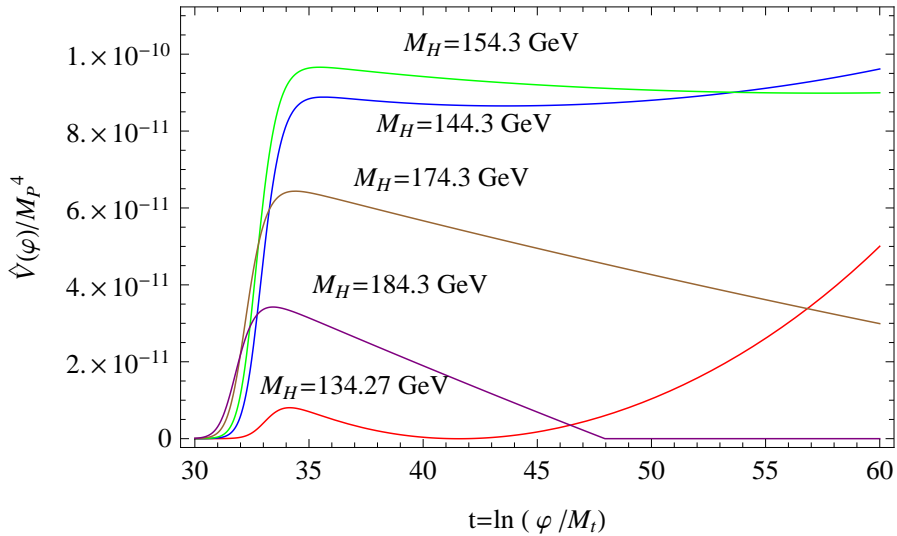}}
\caption{\small The succession of effective potential graphs above
the EW vacuum instability threshold $M_{\rm H}^{\rm inst}=134.27\ {\rm
GeV}$ up to $M_{\rm H}=184.3$ GeV showing the occurrence of a metastable vacuum
followed for high $M_{\rm H}$ by the formation of a negative slope branch.
Local peaks of $\hat V$ situated at $t=34\div35$ grow with $M_{\rm H}$ for
$M_{\rm H}\lesssim 160$ GeV and start decreasing for larger $M_{\rm
  H}$ \cite{RGH}.
 \label{Fig.1}}
\end{figure}

Now we want to show that, in addition to the good agreement of the
spectrum of cosmological perturbations with the CMB data, this model
can also describe the mechanism of generating the cosmological {\em
background} itself upon which these perturbations exist. This
mechanism consists in the formation of the initial conditions for
inflation in the form of a sharp probability peak in the
distribution function (\ref{partitionphi}) at some appropriate value
of the inflaton field $\varphi_0$ with which the Universe as a whole
starts its evolution. The shape and the magnitude of the potential
(\ref{hatV}) depicted in Fig.1 for several values of the Higgs mass
clearly indicates the existence of such a peak.

Indeed, the negative of the inverse potential
damps to zero after exponentiation
the probability of those values of $\varphi$ at which
$\hat V(\varphi)=0$ and, vice versa, enhances the probability at the
positive maxima of the potential. The pattern of this behavior with
growing Higgs mass $M_{\rm H}$ is as follows.

\begin{figure}[h]
\centerline{\epsfxsize 12cm \epsfbox{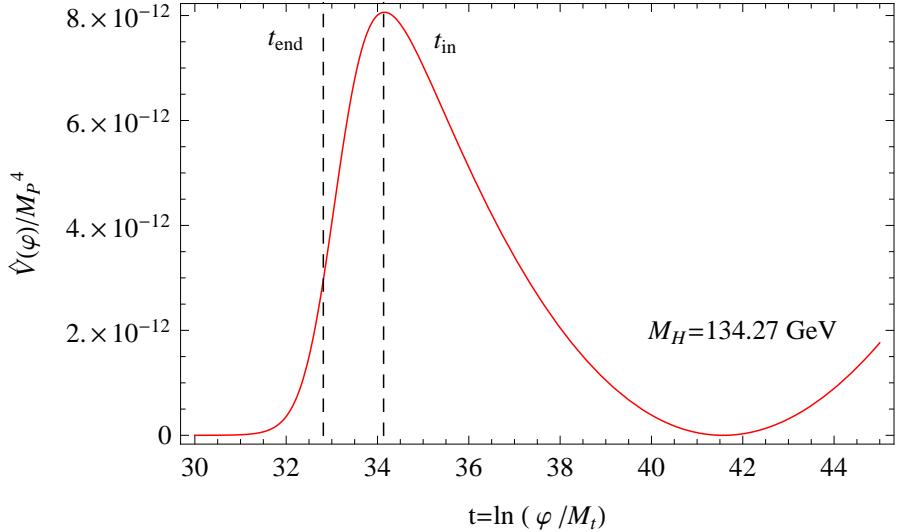}}
\caption{\small The effective potential for the instability
threshold $M_{\rm H}^{\rm inst}=134.27$ GeV. A false vacuum occurs at the
instability scale $t_{\rm inst}\simeq 41.6$, $\varphi\sim 80 M_{\rm P}$.
An inflationary domain for a $N=60$ CMB perturbation (ruled out by the
WMAP bounds) is marked by dashed lines \cite{RGH}. \label{Fig.5}}
\end{figure}

As is known, for low $M_{\rm H}$ the SM has a domain of unstable EW
vacuum, characterized by negative values of running $\lambda(t)$ at
certain energy scales. Thus we begin with the EW vacuum instability
threshold \cite{espinosa,Sher} which exists in this gravitating SM
at $M_{\rm H}^{\rm inst}\approx 134.27$ GeV \cite{RGH,PLBRGH} and which is
slightly lower than the CMB compatible range of the Higgs mass
($M_{\rm H}^{\rm inst}$ is chosen in Fig.~2 and for the lowest curve
in Fig.~1). The potential $\hat V(\varphi)$
drops to zero at $t_{\rm inst}\simeq 41.6$, $\varphi_{\rm inst}\sim
80 M_{\rm P}$, and forms a false vacuum \cite{RGH,PLBRGH} separated from
the EW vacuum by a large peak at $t\simeq 34$. Correspondingly, the
probability of creation of the Universe with the initial value of
the inflaton field at the EW scale $\varphi=v$ and at the instability
scale $\varphi_{\rm inst}$ is damped to zero, while the most
probable value belongs to this peak. The inflationary stage of the
formation of the pivotal $N=60$ CMB perturbation (from the moment
$t_{\rm in}$ of the first horizon crossing until the end of inflation
$t_{\rm end}$), which is marked by dashed lines in Fig.2, lies to
the left of this peak. This conforms to the requirement of the
chronological succession of the initial conditions for inflation and
the formation of the CMB spectra.

\begin{figure}[h]
\centerline{\epsfxsize 12cm \epsfbox{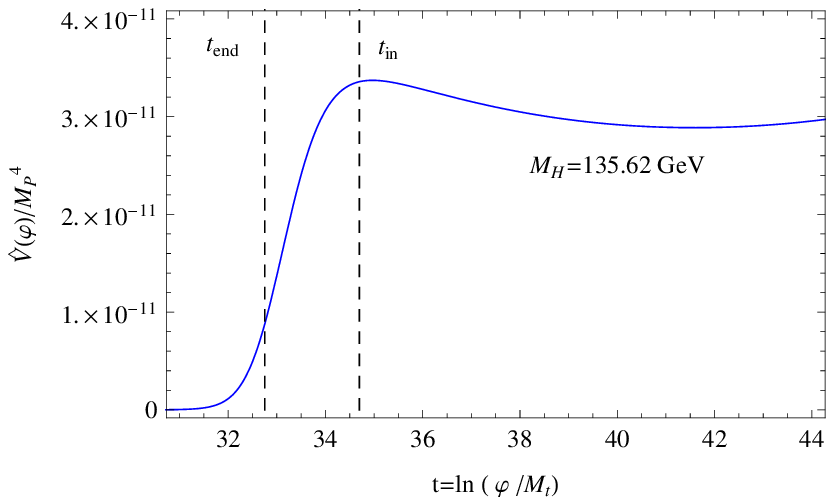}}
\caption{\small Inflaton potential at the lowest CMB compatible
value of $M_{\rm H}$ with a metastable vacuum at $t\simeq 42$ \cite{RGH}.
 \label{Fig.3}}
\end{figure}

The above case is, however, below the CMB-compatible range of $M_{\rm H}$
and was presented here only for illustrative purposes\footnote{Another
interesting range of $M_{\rm H}$ is below the instability threshold
$M_{\rm H}^{\rm inst}$ where $\hat V$ becomes negative in the ``true'' high
energy vacuum. As mentioned in the previous section, the tunneling
state rules out aperiodic solutions of effective equations with
$H^2<0$, which cannot contribute to the quantum ensemble of
expanding Lorentzian signature models. Therefore, this range is
semiclassically ruled out not only by the instability arguments, but
also contradicts the tunneling prescription.}. An important situation
occurs
at higher Higgs masses from the lower CMB bound on $M_{\rm H}\simeq
135.6$ GeV until about 160 GeV. Here we get a family of a metastable
vacua with $\hat V>0$. An example is the plot for the lower CMB
bound $M_{\rm H}=135.62$ GeV depicted in Fig. \ref{Fig.3}. Despite the
shallowness of this vacuum its small maximum generates via
(\ref{partitionphi}) a sharp probability peak for the initial
inflaton field. This follows from an extremely small value of $\hat
V/M_{\rm P}^4\sim 10^{-11}$, the reciprocal of which generates a rapidly
changing exponential of (\ref{partitionphi}). The location of the
peak again precedes the inflationary stage for a pivotal $N=60$ CMB
perturbation (also marked by dashed lines in Fig. \ref{Fig.3}).

For even larger $M_{\rm H}$ these metastable vacua get replaced by a
negative slope of the potential which interminably decreases to zero
at large $t$ (at least within the perturbation theory range of the
model), see Fig.~\ref{Fig.1}. Therefore, for large $M_{\rm H}$ close to
the upper CMB bound 185 GeV, the probability peak of
(\ref{partitionphi}) gets separated from the non-perturbative domain
of large over-Planckian scales due to a fast drop of $\hat
V\sim\lambda/\xi^2$ to zero. This, in turn, follows from the fact
that $\xi(t)$ grows much faster than $\lambda(t)$ when they both
start approaching their Landau pole \cite{RGH}.

The location $\varphi_0$ of the probability peak and its quantum
width can be found in analytical form, and their derivation
shows the crucial role of the running $\mbox{\boldmath$A_I$}(t)$ for
the formation of initial conditions for inflation. Indeed, the
exponential of the tunneling distribution (\ref{partitionphi}) for
$M_{\rm P}^2/\xi\varphi^2\ll 1$ reads as
    \begin{eqnarray}
    \varGamma_-(\varphi)=24\pi^2\frac{M_{\rm P}^4}{\hat V(\varphi)}
    \simeq 96\pi^2\frac{\xi^2}
    \lambda\left(1+\frac{2M_{\rm P}^2}{\xi Z^2\varphi^2}\right),
    \end{eqnarray}
and in view of the RG equations (\ref{beta-lambda})--(\ref{beta-xi})
has an extremum satisfying the equation
    \begin{eqnarray}
    \varphi\frac{d\varGamma}{d\varphi}=\frac{d\varGamma}{dt}
    =-\frac{6\xi^2}
    \lambda\left(\mbox{\boldmath$A_I$}
    +\frac{64\pi^2M_{\rm P}^2}{\xi Z^2\varphi^2}\right)=0,
    \end{eqnarray}
where we again neglect higher order terms in $M_{\rm P}^2/\xi
Z^2\varphi^2$ and $\mbox{\boldmath$A_I$}/64\pi^2$ (extending beyond
the one-loop order). Here, $\mbox{\boldmath$A_I$}$ is the anomalous
scaling  introduced in (\ref{Aviamasses}) and (\ref{AI}) -- the
quantity that should be negative for the existence of the solution
for the probability peak,
    \begin{eqnarray}
    \varphi^2_0=
    \left.-\frac{64\pi^2M_{\rm P}^2}{\xi \mbox{\boldmath$A_I$}Z^2}
    \,\right|_{\;t=t_0}.                               \label{root}
    \end{eqnarray}
As shown in \cite{RGH,PLBRGH}, this quantity is indeed negative. In
the CMB-compatible range of $M_{\rm H}$ its running starts from the range
$-36\lesssim\mbox{\boldmath$A_I$}(0)\lesssim-23$ at the EW scale and
reaches small but still negative values in the range
$-11\lesssim\mbox{\boldmath$A_I$}(t_{\rm end})\lesssim -2$ at the
inflation scale. Also, the running of $\mbox{\boldmath$A_I$}(t)$ and
$Z(t)$ is very slow -- the quantities belonging to the two-loop
order -- and the duration of inflation is very short $t_0\sim t_{\rm
in}\simeq t_{\rm end}+2$ \cite{RGH,PLBRGH}. Therefore,
$\mbox{\boldmath$A_I$}(t_{\rm 0})\simeq\mbox{\boldmath$A_I$}(t_{\rm
end})$, and these estimates apply also to
$\mbox{\boldmath$A_I$}(t_{\rm 0})$. As a result, the second
derivative of the tunneling on-shell action is positive and very
large,
    \begin{eqnarray}
    \frac{d^2\varGamma_-}{dt^2}
    \simeq
    -\frac{12\xi^2}\lambda \mbox{\boldmath$A_I$}\gg 1,
    \end{eqnarray}
which gives an extremely small value of the quantum width of the
probability peak,
    \begin{eqnarray}
    \frac{\Delta\varphi^2}{\varphi^2_0}
    =-\left.\frac\lambda{12\xi^2}
    \frac1{\mbox{\boldmath$A_I$}}\right|_{\,t=t_0}\sim
    10^{-10}.
    \end{eqnarray}
This width is about $(24\pi^2/|{\mbox{\boldmath$A_I$}}|)^{1/2}$
times -- one order of magnitude -- higher than the CMB perturbation
at the pivotal wavelength $k^{-1}=500$ Mpc (which we choose to
correspond to $N=60$). The point $\varphi_{\rm in}$ of the horizon
crossing of this perturbation (and other CMB waves with different
$N$'s) easily follows from equation (\ref{xversusvarphi}) which in
view of $\mbox{\boldmath$A_I$}(t_{\rm
0})\simeq\mbox{\boldmath$A_I$}(t_{\rm end})$ takes the form
    \begin{eqnarray}
    &&\frac{\varphi^2_{\rm in}}{\varphi_0^2}=
    1-\exp\left(-N\frac{|\mbox{\boldmath$A_I$}
    (t_{\rm end})|}{48\pi^2}\right).            \label{phi0phiin}
    \end{eqnarray}
It indicates that for wavelengths longer than the pivotal one the
instant of horizon crossing approaches the moment of ``creation'' of
the Universe, but always stays chronologically succeeding
it, as it must.

%%%%%%%%%%%%%%%%%%%%%%%%%%%%%%%%%%%%%%%%%%%%%%%%%%%%%%%%%%%%%%%%%%%%
\section{Conclusions and discussion}

In this paper we have constructed the tunneling quantum state of the Universe
based on the path integral for the microcanonical ensemble in
cosmology.  The corresponding apparent ensemble
from the quantum state exists in the
unbounded positive range of the effective cosmological constant,
unlike the no-boundary state discussed in \cite{slih,why}
whose apparent ensemble is bounded by the
reciprocated coefficient of the topological term in the overall
conformal anomaly. Also, in contrast to the no-boundary case, the
tunneling state turns out to be a radiation-free vacuum one.

The status of the tunneling versus no-boundary states is rather
involved. In fact, the formal Euclidean path integral (\ref{1}) is a
transformed version of the microcanonical path integral over
Lorentzian metrics, so that its lapse function integration runs
along the imaginary axis from $-{\rm i}\infty$ to $+{\rm i}\infty$
\cite{why}\footnote{This might seem being equivalent to the
tunneling path integral of \cite{Vilenkin84,Vilenkin94}, but the
class of metrics integrated over is very different. We do not impose
by hands $a_-=0$ as the boundary condition, but derive it from the
saddle-point approximation for the integral over formally periodic
configurations. The fact that periodicity gets violated by the
boundary condition $a_-=0$ implies that the a priori postulated
tunneling statistical ensemble is exhausted at the dynamical level
by the contribution of a pure vacuum state \cite{slih,why}.}. The
absence of periodic solutions for stationary points of (\ref{1})
with the Lorentzian signature makes one to distort the contour of
integration over $N$ into a complex plane, so that it traverses the
real axis at the points $N=+1$ or $N=-1$ which give rise to
no-boundary or tunneling states. One can show that the no-boundary
thermal part of the statistical sum of \cite{slih} is not analytic
in the full complex plane of $N$. The $N\gtrless 0$ domains are
separated by the infinite sequence of its poles densely filling the
imaginary axes of $N$. Therefore, the contour of integration passing
through both points $N=\pm1$ is impossible, and the no-boundary and
tunneling states cannot be obtained by analytic continuation from
one another\footnote{In the case of the vacuum no-boundary state
when the vanishing thermal part of the effective action cannot
present an obstacle to analytic continuation in the complex plane of
$N$ the situation stays the same. Indeed, any integration contour
from $-i\infty$ to $+i\infty$ crosses the real axes an odd number of
times, so that the contribution of only one such crossing survives,
because any two (gauge-equivalent) saddle points traversed in
opposite directions give contributions canceling one another.}. They
represent alternative solutions (quantum states)
of the Wheeler-DeWitt equation.

The path-integral formulation of the tunneling state admits a
consistent renormalization scheme and a RG resummation which is very
efficient in cosmology according to a series of recent papers
\cite{Wil,BezShap1,BezShap3,RGH,PLBRGH,Clarcketal}. For this reason
we have applied the obtained tunneling distribution to a recently
considered model of inflation driven by the SM Higgs boson
non-minimally coupled to curvature. In this way a complete
cosmological scenario was obtained, embracing the formation of
initial conditions for the inflationary background (in the form of a
sharp probability peak in the inflaton field distribution) and the
ongoing generation of the CMB perturbations on this background. As
was shown in \cite{RGH,PLBRGH}, the comparison of the CMB amplitude
and the spectral index with the WMAP observations impose bounds on
the allowed range of the Higgs mass. These bounds turn out to be
remarkably consistent with the widely accepted EW vacuum stability
and perturbation theory restrictions. Interestingly, the behavior of
the running anomalous scaling $\mbox{\boldmath$A_I$}(t)<0$, being
crucially important for these bounds, also guarantees the existence
of the obtained probability peak. The quantum width of this peak is
one order of magnitude higher than the amplitude of the CMB spectrum
at the pivotal wavelength, which could entail interesting
observational consequences. Unfortunately, this quantum width is
hardly measurable directly because it corresponds to an infinite
wavelength perturbation (a formal limit of $N\to\infty$ in
(\ref{phi0phiin})), but indirect effects of this quantum trembling
of the cosmological background deserve further study.

We have entertained here the idea that we can obtain sensible
predictions from peaks in the cosmological wavefunction. This is, of
course, different from approaches based on the anthropic principle. We
find it intriguing, however, that a consistent scenario based on our
more traditional approach may be possible and even falsifiable.

To summarize, the obtained results bring to life a convincing
unification of quantum cosmology with the particle phenomenology of the
SM, inflation theory, and CMB observations. They support the
hypothesis that an appropriately extended Standard Model
\cite{nuMSM,dark} can be a consistent quantum field theory all the
way up to quantum gravity and perhaps explain the fundamentals of
all major phenomena in early and late cosmology.

%%%%%%%%%%%%%%%%%%%%%%%%%%%%%%%%%%%%%%%%%%%%%%%%%%%%%%%%%%%%%%%%%%

\section*{Acknowledgements}
The authors express their gratitude to A.A.Starobinsky for fruitful
and thought provoking discussions. A.B. and A.K. acknowledge support by
the grant 436 RUS 17/3/07 of the German Science Foundation (DFG) for
their visit to the University of Cologne. The work of A.B. was also
supported by the RFBR grant 08-02-00725 and the grant
LSS-1615.2008.2. A.K. was partially supported by the RFBR grant
08-02-00923, the grant LSS-4899.2008.2 and by the Research Programme
``Elementary Particles'' of the Russian Academy of Sciences. The work
of C.F.S. was supported by the Villigst Foundation. A.B.
acknowledges the hospitality of LMPT at the University of Tours.

\setcounter{equation}{0}
\renewcommand{\theequation}{A.\arabic{equation}}

%%%%%%%%%%%%%%%%%%%%%%%%%%%%%%%%%%%%%%%%%%%%%%%%%%%%%%%%%%%%%%%%%%%%%

\end{document}